\documentclass[a4paper,fleqn,usenatbib]{mnras}
\usepackage{newtxtext,newtxmath}

\usepackage[T1]{fontenc}
\usepackage{ae,aecompl}
\usepackage{longtable}
\usepackage{graphicx}	
\usepackage{amsmath}	
\usepackage{amssymb}	

\title[SN 2018aoq]{Supernova 2018aoq and a distance to Seyfert galaxy 
NGC\,4151}

\author[D. Yu. Tsvetkov et al.]{
D. Yu. Tsvetkov,$^{1}$\thanks{E-mail: tsvetkov@sai.msu.su}
P. V. Baklanov,$^{2,3,4}$
M. Sh. Potashov,$^{2,3}$
V. L. Oknyansky,$^{1}$
\newauthor
Kh.M. Mikailov, $^{5}$ 
N.A. Huseynov,$^{5}$
I.A. Alekberov,$^{5}$  
O.V. Khalilov,$^{5}$ 
\newauthor
N. N. Pavlyuk,$^{1}$ 
V. G. Metlov,$^{1}$
I. M. Volkov$^{1,6}$
and S. Yu. Shugarov$^{1,7}$
\\
$^{1}$Sternberg Astronomical Institute, M.V. Lomonosov Moscow 
State University, Universitetsky pr. 13, Moscow 119234, Russia\\
$^{2}$Institute for Theoretical and Experimental Physics,
ul. Bolshaya Cheremushkinskaya 25, Moscow 117259, Russia\\
$^{3}$Novosibirsk State University, ul. Pirogova 2, Novosibirsk, Russia\\
$^{4}$National Research Nuclear University MEPhI, Kashirskoe sh. 31, Moscow
115409, Russia\\ 
$^{5}$Shamakhy Astrophysical Observatory, National Academy of Sciences, 
AZ 5626 Pirkuli, Azerbaijan\\
$^{6}$ Institute of Astronomy of the Russian Academy of Sciences, 
48 Pyatnitskaya street, 119017 Moscow \\
$^{7}$ Astronomical Institute of the Slovak Academy of Sciences, 059 60 
Tatransk\'a Lomnica, The Slovak Republic\\ 
}

\date{Accepted XXX. Received YYY; in original form ZZZ}

\pubyear{2019}

\begin{document}
\label{firstpage}
\pagerange{\pageref{firstpage}--\pageref{lastpage}}
\maketitle

\begin{abstract}
We present optical photometric observations
of SN\,2018aoq from 2 to 100 days after explosion,
and 7 spectra at epochs from 11 to 71 days.
The light curves and spectra are typical for SNe II-P.
As previously reported,
SN\,2018aoq appears to be of intermediate brightness between
subluminous and normal SNe II-P.
SN\,2018aoq was discovered in Seyfert galaxy NGC\,4151, for which
the distance is uncertain.
We utilised the Expanding Photosphere Method using three
sets of filter combinations 
and velocities derived from the absorption minima
of FeII lines and obtained    
a distance of  
20.0$\pm$1.6 Mpc.
The Standard Candle Method applied to SN\,2018aoq yields 
a distance of 16.6$\pm$1.1 Mpc.
Both values are consistent with the distance measurements for 
NGC\,4151 based on geometric method.
\end{abstract}




\section{Introduction}

NGC\,4151 is well-known Seyfert 1 galaxy, 
one of the nearest galaxies with active nucleus.

NGC\,4151 is one of the rare objects for which there exist two 
independent dynamical measurements for the mass of the central black hole.
To first order, the black hole mass derived by the stellar  
dynamical modeling depends linearly on the assumed distance
to the galaxy \citep{onken2014}.

However, the actual distance to the galaxy is rather  
uncertain.

The Extragalactic Distance Database \citep{tully2009}
presents distance measurements based on the Tully-Fisher relation:
the individual estimate for NGC\,4151 is
3.9$\pm$0.4 Mpc, and the group-average distance is 11.2$\pm$1.1 Mpc.
The reability of these distance estimates are doubtful, as
discussed by \citet{onken2014}.

The methods based on the reprocessing of the emission of the
active nucleus provided much larger distances: 19 Mpc \citep{cackett2007} 
and 29 Mpc \citep{yoshii2014}. 
\citet{honig2014} applied a geometric method, measuring the size
of region of hot dust emission as determined from time-delays and
infrared interferometry, which yielded 19.0$\pm$2.5 Mpc.

The discovery of a type II-P supernova (SN) 2018aoq in NGC\,4151 presents
a new possibility to obtain an independent estimate of the distance to
the galaxy. 

The optical transient Kait-18P=2018aoq was discovered on 
2018-04-01.4316 by the Lick Observatory Supernova Search at the 
unfiltered magnitude of 15.3 at a distance of
73\arcsec\ from the center of NGC\,4151. 
Spectroscopic observations with the 1.5-m Kanata telescope 
classified the transient as a Type II supernova\footnotemark.
\footnotetext{https://wis-tns.weizmann.ac.il/search}

Observations of type II-P SNe can be used to determine distances 
to their host galaxies using the Expanding Photosphere Method (EPM),
which was first developed by \citet{Kirshner1974}. 
The method is based on measuring the angular radius of the 
photosphere from photometric data and comparing the resulting
expansion rate to the velocity extracted from the spectral data.
The EPM provides
estimates of distance independent of extragalactic distance ladder.
The method requires high-quality spectroscopic and photometric
monitoring of SNe and was applied mostly to nearby objects
\citep[e.g.,][]{Hamuy2001, Takats2006, Jones2009, 
Bose2014}, 
although recently it became possible to perform the EPM on SNe
at cosmologically significant redshifts
\citep[e.g.,][]{Gall2016, Gall2018}.
The other method for distance determinations using SNe\,II-P is the
Standardized Candle Method (SCM) \citep{Hamuy2002}, based on a 
correlation between the luminosity and the expansion velocity
of SNe during the plateau phase. This method relies on the local distance
calibrators and yields distances that are in reasonable agreement 
with the EPM \citep[e.g.,][]{Nugent2006, Poznanski2009, Olivares2010,
Gall2018}. 
   
\section {Observations}

Photometric $UBVRI$ CCD observations of SN\,2018aoq were carried out at the 
60-cm and 50-cm telescopes of 
Crimean Observatory of
Sternberg Astronomical Institute (SAI), the 70-cm and 20-cm telescopes
of Moscow Observatory of SAI,
the 1-m telescope of Institute of Astronomy of Russian Academy
of Science (INASAN) at 
Simeiz Observatory, 
the 60-cm telescope of Star\'a
Lesn\'a Observatory of the Astronomical Institute of Slovak Academy of
Science, and the 60-cm telescope of Shamakhy Astrophysical Observatory. 

The standard image reductions and photometry were made using the
{\sc IRAF}\footnotemark .
\footnotetext{{\sc IRAF} is distributed by the National Optical
Astronomy Observatory,
which is operated by AURA under cooperative agreement with the
National Science Foundation.}
The magnitudes of the SN
were derived by a PSF-fitting relatively to
a sequence of local standard stars, which were calibrated by 
\citet{lyutyi1973}, 
\citet{doroshenko2005}, and
\citet{roberts2012}.
The photometry was transformed to standard Johnson-Cousins
$UBVRI$ magnitudes by means of instrumental colour-terms. 

The surface brightness of the host galaxy at the location of the SN
is not very high, nevertheless we checked if the galaxy background affects
the photometry. We used the images obtained before SN outburst at the
Shamakhy Observatory for galaxy subtraction. We found that for most of
the images the effect of galaxy background does not exceed the errors
of magnitudes, but for the 50-cm telescope of SAI it may amount to 0.05-0.1 mag.
We applied galaxy subtraction for all images obtained with this telescope.

The magnitudes of standard stars are presented in Table~
\ref{tab:localstand}, 
the photometric data are presented in Table~\ref{tab:photometry}. 

\begin{figure}
\includegraphics[width=\columnwidth]{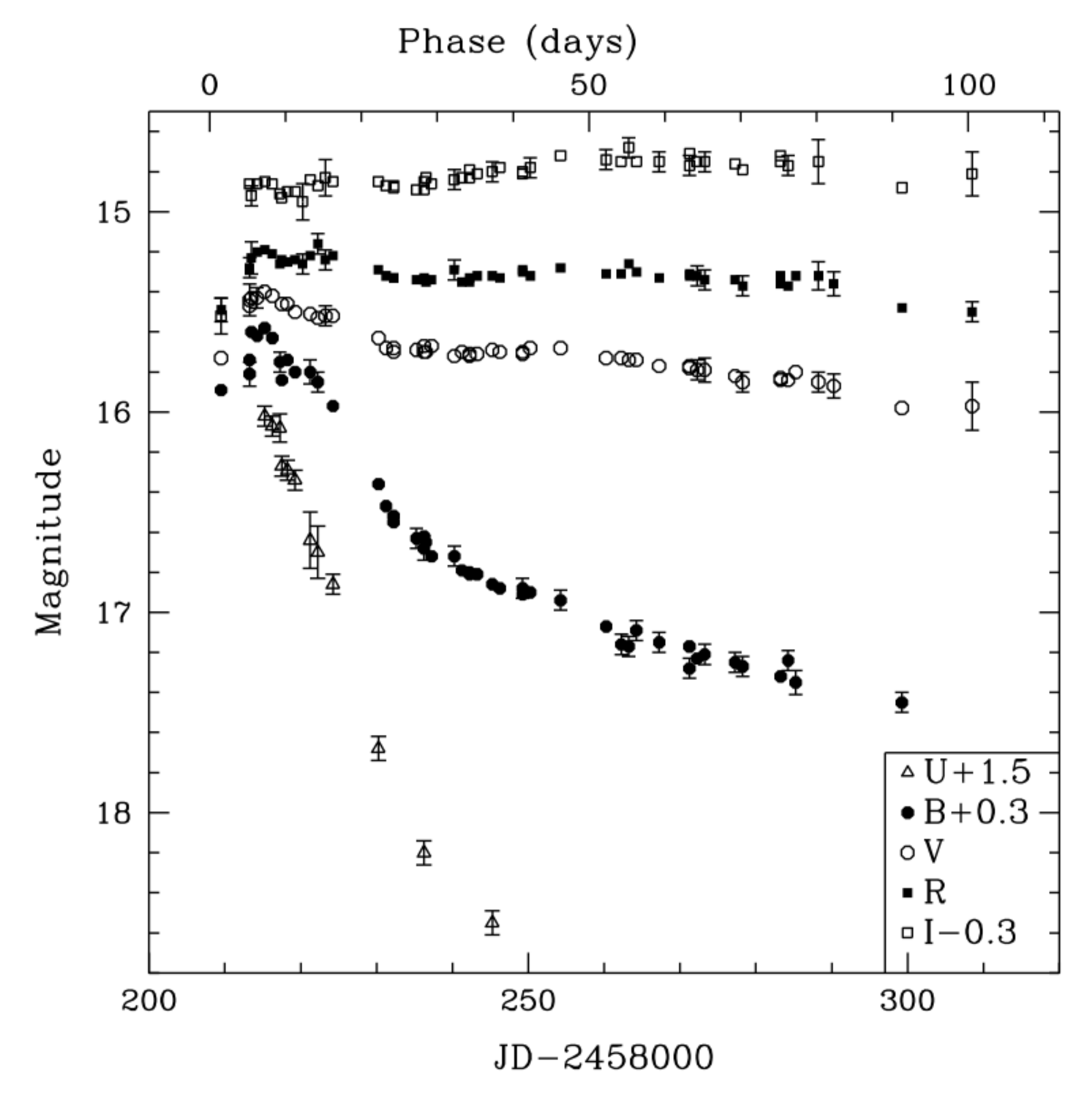}
\caption{The light curves of SN\,2018aoq. The phase is in reference
to the explosion date JD\,2458208. The error bars are shown only if
they exceed the size of a symbol}
    \label{fig:ligcur}
\end{figure}

Prediscovery observations were reported by \citet{nazarov2018}.
We carried out photometry on their images, using our local standard
stars and applying galaxy subtraction, and obtained new magnitude estimates, 
which are also reported in Table~\ref{tab:photometry}.  
The light curves are shown in Fig.~\ref{fig:ligcur}

The shape of the light curves is typical for SNe II-P. 
The first observations were obtained on the rising part of the light
curves,
and we can determine the epoch when the SN reached the plateau phase
as JD\,2458215$\pm$1 (April 6).  
\citet{yamanaka2018} obtained images of NGC\,4151 on JD\,2458209.0
(March 31.5) and derived an upper limit of 17.5 mag in the $R$-band,
\citet{ONeil2019} reported that the SN was fainter than
18.89 mag in the 'orange' ATLAS filter on JD\,2458206.97
(March 29). We used a polynomial fit to the $R$-band magnitudes 
on the rise
and 
found that the best estimate for the epoch of explosion is 
JD\,2458208$\pm$1, 7 days before start of the plateau.
This value of rise time is in agreement with the average
rise time for SNe II-P  
reported by \citet{Gall2015}.

The blue colour of SN\,2018aoq at maximum and the absence of 
detectable interstellar lines in the spectra allows to conclude that 
the absorption in the host galaxy was negligible. The galactic extinction
is small $E(B-V)_{gal}=0.02$ mag \citep{schlafly2011}. 
\citet{ONeil2019}
compared the colour curves of SN\,2018aoq to those of type II-P SNe
for which the extinction is well-known, and derived the total 
extinction for SN\,2018aoq $E(B-V)_{tot}=0.04$ mag, only slightly 
larger than $E(B-V)_{gal}$.    
We used $E(B-V)_{tot}=0.04$ for all further calculations. 

\begin{figure}
\includegraphics[width=\columnwidth]{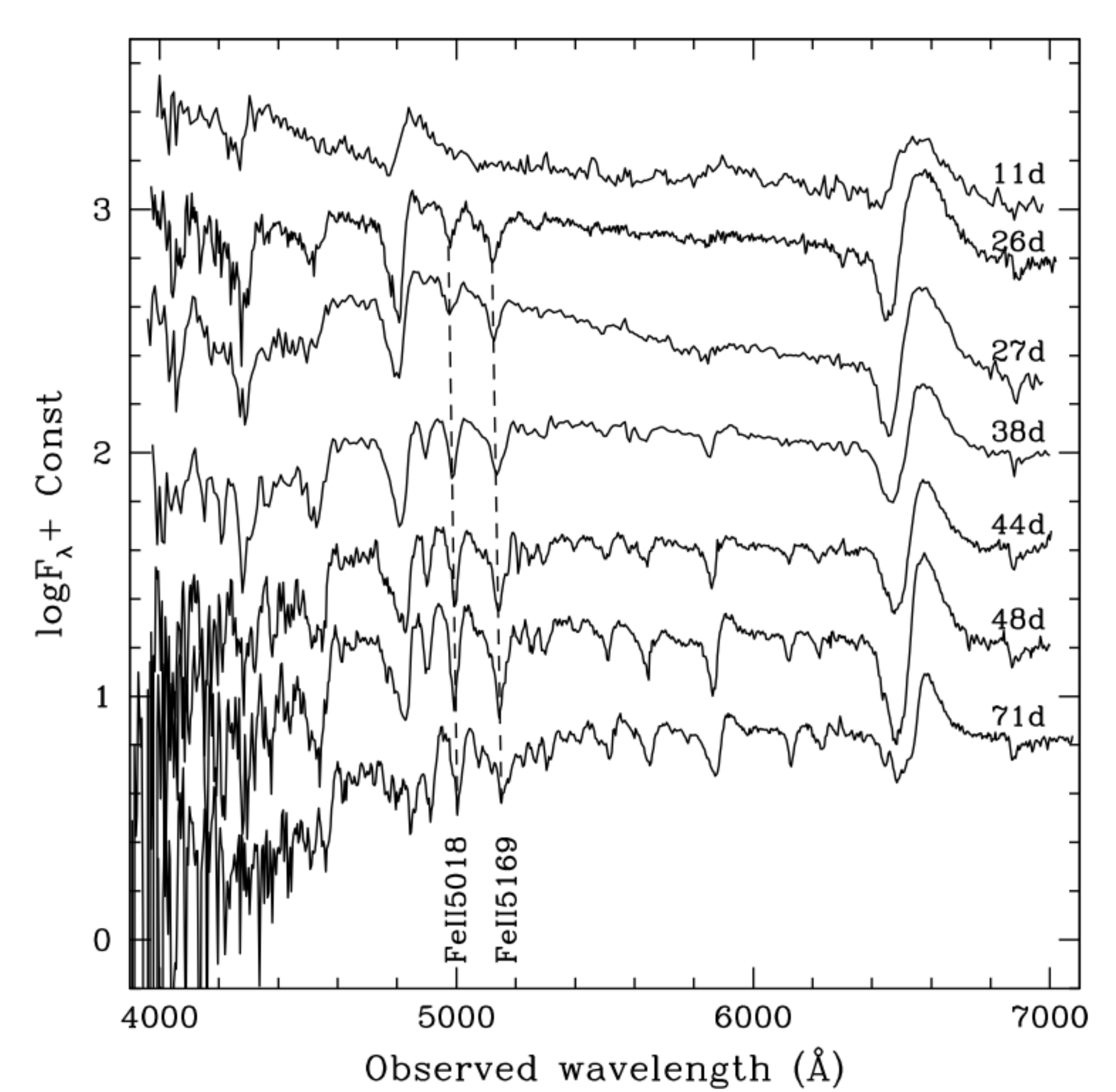}
\caption{The spectra of SN 2018aoq. The ages are relative to 
the date of explosion (JD\,2458208). The vertical dashed lines indicate
absorption minima of FeII lines used for the EPM. These lines are 
not detected in the first spectrum, which was not used for the EPM}
    \label{fig:sp}
\end{figure}

Spectroscopic observations were obtained at the 2-m telescope
of Shamakhy Astrophysical Observatory. The modified Universal Astronomical
Grating Spectrograph provided 
the wavelength range of 3900 -- 7000 \AA\  
with a dispersion of
115\AA\,mm$^{-1}$, which corresponds to 4.1 or 8.2 \AA\,pixel$^{-1}$ for
different CCD binning. The journal of spectroscopic observations is
presented in Table~\ref{tab:spectra}, the spectra are shown 
in Fig.~\ref{fig:sp}.

We continue the observations of SN\,2018aoq, the complete set of data and
its analysis will be presented in a separate paper.

\section{The EPM distance}

The Expanding Photosphere Method (EPM) \citep{Kirshner1974}
determines a distance $D$ to the SN from the 
relation $\theta = R/D$, where $\theta$ is 
the angular 
radius of photosphere, $R$ is its linear radius. 

The method can be applied if the ejecta is spherically symmetric,
the envelope undergoes free expansion, so that
the velocity of matter $v$ and the radial distance $r$ are connected by
$v = r/(t-t_0)$, where $t_0$ is 
the zero-point time, which might be offset from
the true moment of explosion.
 
The photospheric flux of SN is described by a modified
Planck spectrum $F_\nu(R) = \zeta^2 \pi B_\nu(T_{\rm col})$, 
where $\zeta$
is the correction factor, 
$T_{\rm col}$ is the colour temperature,
$B_\nu(T_{\rm col})$ is the Planck function.

The description of the EPM is presented in a number of papers,
\citep[e.g.,][]{Hamuy2001, Takats2006, Jones2009, Gall2018}. 
We applied the EPM following the prescriptions of \citet{Hamuy2001}. 

The correction factor $\zeta$ cannot be determined from observations.
The empirical relations between $\zeta$ and $T_{\rm col}$ were 
established by \citet{Eastman1996} and \citet{Dessart2005}.
We used the relation by \citet{Dessart2005}, which is  
confirmed by our research (Baklanov, in prep.) 
and by \citet{Vogl2018}.

We used three sets of filter combinations to derive the temperature and 
angular radius of SN photosphere.  
The errors in quantities $\theta$ and $T_{\rm col}$ were estimated
using Monte Carlo technique. Samples of data points were drawn 
from normal distributions of uncertainty in the photometric
fluxes.

The velocity of matter at the photosphere $v_{\rm ph}$ can be measured by 
the blueshift
of weak absorption lines, the lines of FeII $\lambda$5018\AA\  and 
$\lambda$5169\AA\  are used
more often \citep{Takats2012}. 
The observed spectra 
were corrected for the redshift of the galaxy $z=0.00332$\footnotemark
\footnotetext{https://ned.ipac.caltech.edu/}
and continuum subtracted, using polynome fitting with 
the {\sc SNID} package \citep{Blondin2007}.
The spectra were smoothed with a Savitzky-Golay filter \citep{Savitzky1964}
and wavelengths of
absorption minima were
determined.
The uncertainties of velocity measurements were estimated to 
be in the range of 4--6\%, depending on the spectral resolution
and photon statistics of the detector.  

The computations were carried out for three sets of filter combinations:
$BVI$, $BV$, and $VI$, for the velocities derived from 
the lines FeII$\lambda$5018, $\lambda$5169 and
for the average velocities. 

Table ~\ref{tab:epm} presents the basic EPM quantities. 
$T_{\rm col}$,  $\zeta$ and $\theta$ are given only for the $BVI$ 
filter set, for other sets they are similar.

\begin{table*}
\begin{center}
\caption{The EPM quantities derived for SN\,2018aoq. 
The uncertainties are in parentheses.}
\label{tab:epm}
\begin{tabular}{ccccccc} 
\hline
JD &Phase,& $v$(FeII$\lambda$5018) & $v$(FeII$\lambda$5169) & 
  $T_{\rm col}(BVI) $ & $\zeta$ & $\theta$  \\
        2458200+ &days&   km\,s$^{-1}$  &  km\,s$^{-1}$ &
           K     &   &         $10^{-11}$ rad \\  	
\hline
34.36 &26.4& 3529 (140) & 3893 (154)  & 7022 (225) & 0.662 & 1.62 (0.11) \\
35.33 &27.3& 3529 (234) & 3893 (231)  & 6968 (212) & 0.667 & 1.63 (0.10) \\ 
46.35 &38.4& 3125 (185) & 3083 (183)  & 6245 (181) & 0.747 & 1.80 (0.12) \\
52.45 &45.5& 2707 (108) & 2680 (106)  & 5975 (184) & 0.786 & 1.88 (0.14) \\
56.40 &48.4& 2706 (108) & 2666 (106)  & 5703 (187) & 0.833 & 1.95 (0.17) \\
79.39 &71.4& 1903 (75) &  1862 (90)   & 5365 (135) & 0.905 & 1.94 (0.13) \\
\hline
\end{tabular}
\end{center}
\end{table*}

\begin{figure}
\includegraphics[width=\columnwidth]{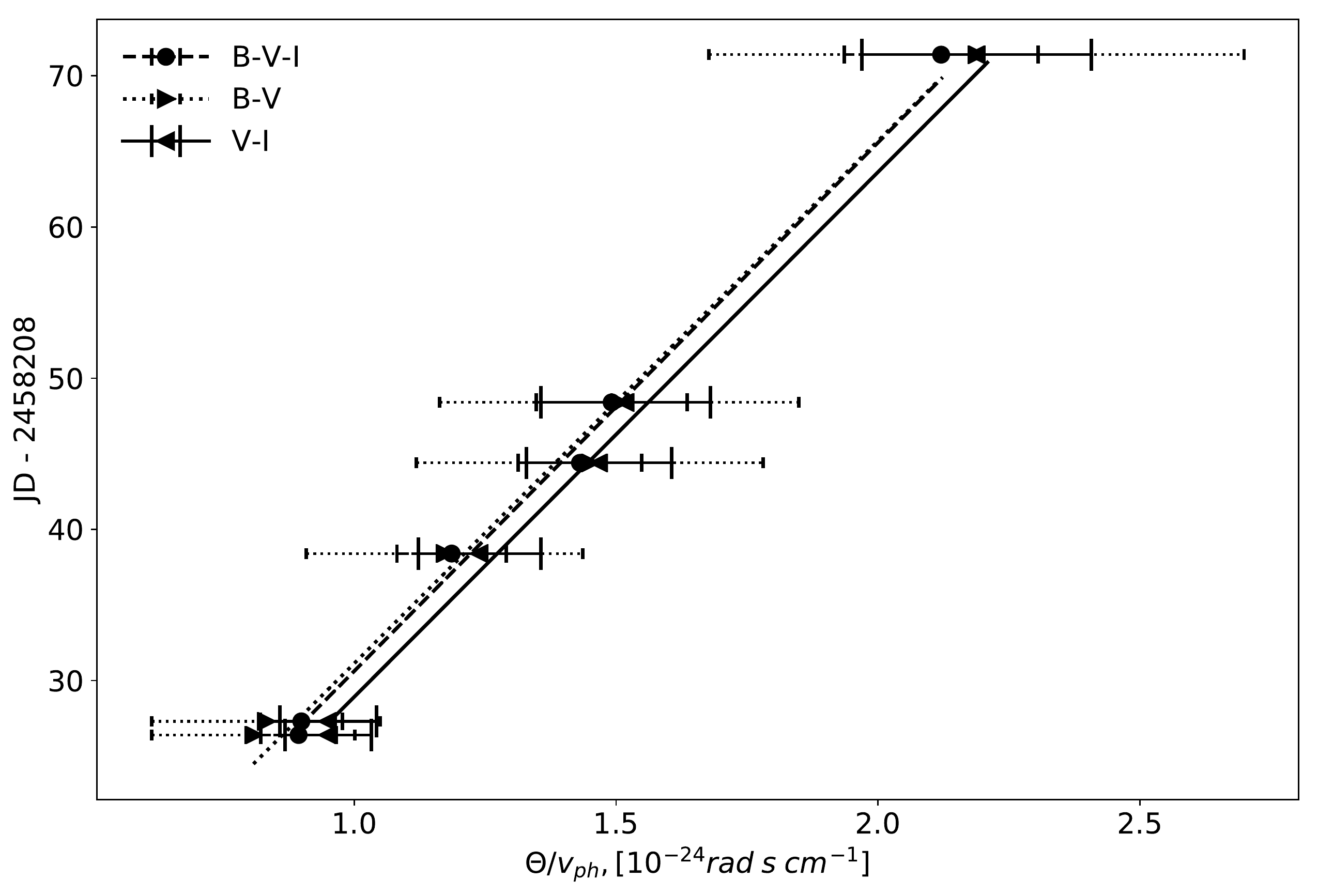}
\caption{The ratio
$\theta/v$ as a function of time for three filter sets,
for average velocity.
}
\label{fig:epm}
\end{figure}

The ratio $\theta/v$ as a function of time for
the filter sets $BVI$, $BV$, and $VI$
is presented in Fig.~\ref{fig:epm}.

We determined $t_0$ and $D$ using the Markov Chain Monte Carlo method 
in the {\sc EMCEE} software package \citep{Foreman-Mackey2012}.

The results are presented in
Table~\ref{tab:dist}.  

\begin{table}
\begin{center}
\caption{The EPM distances for SN\,2018aoq. 
The uncertainties are in parentheses.}
\label{tab:dist}
\begin{tabular}{cccc} 
\hline
 Filter set & FeII line  & $D$, Mpc & $t_0$, JD\,2458000+ \\ 
\hline
$BVI$ & 5018    & 21.4 (2.6) & 200.3 (5.5) \\
$BVI$ & 5169    & 19.4 (1.7) & 205.8 (3.2) \\
$BVI$ & Average & 20.2 (2.1) & 203.5 (4.3) \\
$BV$  & 5018    & 19.9 (3.9) & 203.3 (7.7) \\
$BV$  & 5169    & 19.7 (3.6) & 204.3 (6.5) \\
$BV$  & Average & 20.0 (3.9) & 203.3 (7.6) \\
$VI$  & 5018    & 21.0 (3.3) & 199.2 (7.0) \\
$VI$  & 5169    & 19.1 (2.2) & 204.6 (4.4) \\
$VI$  & Average & 19.8 (2.7) & 202.4 (5.6) \\
\hline
\end{tabular}
\end{center}
\end{table}

\section{The SCM distance}

The Standardized Candle Method (SCM) \citep{Hamuy2002} is based on 
a correlation between the
absolute
brightness of SNe II-P and the expansion velocities derived from the 
minimum of the FeII P-Cygni feature 
observed during the plateau phase.

We used our estimates of expansion velocity 
from the shift of 
FeII$\lambda5169$ line and photometry in the $VI$ bands
and applied the SCM using the calibration by \citet{Polshaw2015},
based on the Cepheid distances to well-observed SNe II-P.
We obtain distance estimates
$D_V=16.7\pm1.4$ Mpc, $D_I=16.5\pm1.3$ Mpc, and the average
$D=16.6\pm1.1$ Mpc.

\section{Discussion}

All distance estimates presented in Table~\ref{tab:dist} are
consistent with each other, 
and we may accept the average value $D=20.0\pm1.6$ Mpc
as the EPM distance for SN\,2018aoq and NGC\,4151, which is in 
good agreement with the
result of \citet{honig2014}. 
  
The estimates of $t_0$ are earlier than the explosion epoch
derived from photometry, but for most of the data the difference 
does not exceed the uncertainties.
We should note that the epoch $t_0$ from the EPM fit may be offset from
the explosion date \citep{Takats2006}.  

Recently \citet{ONeil2019} utilised the SCM method for SN\,2018aoq
as calibrated by
\citet{Polshaw2015} to obtain a distance of $18.2\pm1.2$ Mpc.
Our SCM distance is about 9\% shorter   
than the result of \citet{ONeil2019}, because of small differences
in the observational data. 

The expansion velocity of SN 2018aoq is low, about 2600 km s$^{-1}$
at 50 days past explosion. 
The luminosity at the plateau is 
$M_I=-16.4$ mag for the EPM distance, and $M_I=-16.0$
mag for the SCM distance. 
SN\,2018aoq appears to be an intermediate object 
between subluminous and normal SNe II-P, as was suggested by 
\citet{ONeil2019}. 

The distance measurements by the EPM and SCM may have
systematic errors, for the EPM they result  
from the adopted values of 
dilution factor $\zeta$, which may also be a function of chemical 
composition and density structure of the envelope. Other reasons
for uncertainty are the difference of photospheric velocity from that
derived from the FeII lines and absence of spherical symmetry of the 
ejecta. 
The major sources of errors for the SCM are the calibration
process and the internal diversity of the properties of SNe II-P.     
 
In the case of SN\,2018aoq 
the SCM distance is about 18\% shorter than the EPM distance,
but both values are 
consistent with the most
reliable distance estimate for the host galaxy
$D=19.0\pm2.5$ Mpc, based on geometric 
technique \citep{honig2014}.

We may conclude that these results confirm the 
applicability of SNe II-P for 
distance measurements.

\section*{Acknowledgements}

The work of D.Tsvetkov and P.Baklanov was partly supported 
by the Russian Science
Foundation Grant No. 16-12-10519. 
The work of S.Shugarov was partially supported by Grants
VEGA 2/0008/17 and APVV-15-0458.
The work of I.Volkov was supported by the scholarship of 
the Slovak Academic
Information Agency (SAIA), by the 
Russian Science Foundation Grant No. 14-12-00146 and 
Russian Foundation for Basic Research
Grant No. 18-502-12025.
The work on photospheric velocity determination 
was done by M.Sh.Potashov and was supported by 
the Russian Science Foundation Grant No. 19-12-00229.
We thank the anonymous referee for constructive suggestions
which helped to improve the paper.




\bibliographystyle{mnras}
\bibliography{n4151} 


\newpage
\onecolumn
\appendix{A}
\section{Data tables}

\begin{table}
\begin{center}
\caption{$UBVRI$ magnitudes of local standard stars}
\label{tab:localstand}
\begin{tabular}{cccccccc}
\hline
$\alpha$(J2000)&$\delta$(J2000)  & $U$ & $B$ &  $V$ & 
$R$ & $I$ & References\\
\hline
12:09:53.538  & 39:26:00.15 &11.68 & 9.76 & 8.15  &       &       & 1 \\
12:10:31.442  & 39:26:41.26 &13.24 & 12.47& 11.45 & 10.92 & 10.44 & 1,2,3 \\
12:10:25.255  & 39:23:55.45 &      & 15.08& 14.38 & 13.99 & 13.66 & 3 \\
12:10:38.529  & 39:20:29.56 &15.61 & 14.01& 12.77 & 12.10 & 11.50 & 1,2 \\
\hline     
\end{tabular}

1) \citet{lyutyi1973}, 2) \citet{doroshenko2005},
3) \citet{roberts2012}
\end{center} 
\end{table}

\begin{center}
\begin{longtable}{ccccccccccccr}
\caption{$UBVRI$ magnitudes of SN2018aoq}
\label{tab:photometry}\\
\hline
JD& Phase,  & $U$ & $\sigma_U$ & $B$ & $\sigma_B$ & $V$ & $\sigma_V$ &
$R$ & $\sigma_R$ & $I$ & $\sigma_I$& Telescope \\
2458000+ & days & & & & & & & & & & &\\
\hline
\endfirsthead
\caption{Continued.}\\
\hline
JD& Phase,  & $U$ & $\sigma_U$ & $B$ & $\sigma_B$ & $V$ & $\sigma_V$ &
$R$ & $\sigma_R$ & $I$ & $\sigma_I$& Telescope \\
2458000+ & days & & & & & & & & & & &\\
\hline
\endhead
\hline
\endfoot
\hline
\endlastfoot
209.46 &1.5&     &     &  15.59& 0.04 & 15.73& 0.04&  15.49& 0.06&  15.82& 0.09& K70\\
213.30 &5.3&     &     &  15.51& 0.06 & 15.44& 0.08&  15.28& 0.05&       &     & M20\\ 
213.39 &5.4&     &     &  15.44& 0.04 & 15.47& 0.03&  15.29& 0.03&  15.16& 0.04& C50\\ 
213.42 &5.4&     &     &  15.30& 0.04 & 15.43& 0.04&  15.23& 0.08&  15.22& 0.05& C60\\ 
214.39 &6.4&     &     &  15.32& 0.04 & 15.43& 0.05&  15.20& 0.04&  15.16& 0.04& C60\\ 
215.32 &7.3&14.52& 0.05&  15.28& 0.04 & 15.40& 0.04&  15.19& 0.04&  15.15& 0.04& C60\\ 
216.33 &8.3&14.57& 0.05&  15.33& 0.04 & 15.42& 0.03&  15.21& 0.04&  15.16& 0.04& C60\\ 
217.32 &9.3&14.58& 0.07&  15.45& 0.05 &      &     &  15.26& 0.03&  15.21& 0.04& C60\\
217.52 &9.5&14.77& 0.05&  15.54& 0.04 & 15.46& 0.03&  15.24& 0.04&  15.23& 0.04& T60 \\
218.36 &10.4&14.79& 0.05&  15.44& 0.04 & 15.46& 0.04&  15.25& 0.04&  15.20& 0.04& C60\\
219.35 &11.4&14.84& 0.05&  15.50& 0.04 & 15.50& 0.04&  15.24& 0.04&  15.20& 0.04& C60\\
220.27 &12.3&     &     &       &      &      &     &  15.26& 0.05&  15.25& 0.09& M20\\
221.33 &13.3&15.14& 0.14&  15.50& 0.06 & 15.51& 0.03&  15.22& 0.03&  15.14& 0.04& C60\\
222.34 &14.3&15.20& 0.13&  15.55& 0.05 & 15.53& 0.03&  15.16& 0.05&  15.17& 0.03& C60\\
223.28 &15.3&     &     &       &      & 15.52& 0.05&  15.24& 0.05&  15.13& 0.09&  M20\\
224.31 &16.3&15.36& 0.05&  15.67& 0.04 & 15.52& 0.03&  15.22& 0.03&  15.15& 0.03& C60\\
230.36 &22.4&16.18& 0.06&  16.06& 0.04 & 15.63& 0.03&  15.29& 0.03&  15.15& 0.03& C60\\
231.36 &23.4&     &     &  16.17& 0.04 & 15.68& 0.03&  15.32& 0.04&  15.17& 0.03& C50\\
232.30 &24.3&     &     &  16.22& 0.04 & 15.70& 0.04&  15.33& 0.04&  15.17& 0.04& M70\\
232.34 &24.3&     &     &  16.25& 0.04 & 15.68& 0.03&  15.33& 0.04&  15.18& 0.03& C50\\
234.28 &26.3&     &     &       &      & 15.66& 0.03&       &     &  15.21& 0.03& A60\\
235.25 &27.2&     &     &  16.34& 0.07 & 15.67& 0.03&  15.37& 0.03&       &     &A60\\
235.35 &27.4&     &     &  16.33& 0.05 & 15.69& 0.04&  15.34& 0.03&  15.19& 0.04& C50\\
236.29 &28.3&16.70& 0.06&  16.32& 0.04 & 15.67& 0.04&  15.33& 0.03&  15.19& 0.04& T60\\
236.30 &28.3&     &     &  16.38& 0.06 & 15.70& 0.04&  15.34& 0.04&  15.15& 0.04& C50\\
236.47 &28.5&     &     &  16.35& 0.04 & 15.70& 0.03&  15.35& 0.03&  15.13& 0.04& M70\\
237.30 &29.3&     &     &  16.42& 0.04 & 15.67& 0.03&  15.34& 0.03&  15.16& 0.04& C50\\
240.31 &32.3&     &     &  16.42& 0.05 & 15.72& 0.04&  15.29& 0.05&  15.14& 0.05& C50\\
241.26 &33.3&     &     &       &      & 15.76& 0.04&  15.39& 0.03&  15.17& 0.03& A60\\
241.34 &33.3&     &     &  16.49& 0.04 & 15.70& 0.03&  15.35& 0.03&  15.13& 0.03& C50\\
242.28 &34.3&     &     &  16.51& 0.04 & 15.71& 0.04&  15.35& 0.04&  15.09& 0.04& M70\\
242.32 &34.3&     &     &  16.50& 0.04 & 15.72& 0.03&  15.33& 0.03&  15.13& 0.03& C50\\
243.32 &35.3&     &     &  16.51& 0.04 & 15.71& 0.03&  15.32& 0.03&  15.11& 0.03& C50\\
245.34 &37.3&17.05& 0.06&  16.56& 0.04 & 15.69& 0.04&  15.32& 0.04&  15.10& 0.05& C60\\
246.28 &38.3&     &     &  16.58& 0.04 & 15.70& 0.03&  15.33& 0.04&  15.08& 0.04& M70\\
246.41 &38.4&     &     &  16.66& 0.06 & 15.73& 0.04&  15.29& 0.04&  15.11& 0.04&A60\\
249.31 &41.3&     &     &  16.58& 0.05 & 15.71& 0.03&  15.30& 0.04&  15.11& 0.04& M70\\
249.38 &41.4&     &     &  16.61& 0.04 & 15.70& 0.04&  15.29& 0.03&  15.10& 0.04& T60\\
250.32 &42.3&     &     &  16.60& 0.04 & 15.68& 0.03&  15.32& 0.04&  15.08& 0.05& M70\\
253.37 &45.4&     &     &  16.66& 0.04 & 15.74& 0.06&  15.32& 0.05&  15.02& 0.04&A60\\
254.35 &46.4&     &     &  16.64& 0.05 & 15.68& 0.04&  15.28& 0.03&  15.02& 0.04& M70\\
255.42 &47.4&     &     &       &      & 15.72& 0.05&  15.33& 0.05&  15.11& 0.03& A60\\
256.35 &48.4&     &     &  16.60& 0.04 & 15.74& 0.04&  15.33& 0.05&  15.05& 0.04&A60\\
257.36 &49.4&     &     &  16.68& 0.03 & 15.78& 0.04&  15.36& 0.04&  15.03& 0.04&A60\\
260.34 &52.3&     &     &  16.77& 0.04 & 15.73& 0.03&  15.31& 0.04&  15.04& 0.05& M70\\
262.30 &54.3&     &     &  16.86& 0.05 & 15.73& 0.04&  15.31& 0.04&  15.05& 0.04& M70\\
263.33 &55.3&     &     &  16.87& 0.05 & 15.74& 0.04&  15.26& 0.04&  14.98& 0.05& M70\\
264.32 &56.3&     &     &  16.79& 0.05 & 15.74& 0.03&  15.30& 0.03&  15.05& 0.04& M70\\
267.34 &59.3&     &     &  16.85& 0.05 & 15.77& 0.04&  15.33& 0.03&  15.05& 0.05& M70\\
271.31 &63.3&     &     &  16.98& 0.05 & 15.78& 0.04&  15.32& 0.03&  15.07& 0.05& M70\\
271.38 &63.4&     &     &  16.87& 0.04 & 15.77& 0.03&  15.31& 0.03&  15.01& 0.03& S100 \\
272.31 &64.3&     &     &  16.93& 0.04 & 15.79& 0.05&  15.32& 0.05&  15.05& 0.04& S100\\
273.35 &65.4&     &     &  16.91& 0.05 & 15.79& 0.06&  15.34& 0.05&  15.05& 0.05& S100\\
275.27 &67.3&     &     &  16.88& 0.05 & 15.81& 0.07&       &     &  15.02& 0.03&A60\\
277.27 &69.3&     &     &  16.95& 0.05 & 15.82& 0.04&  15.34& 0.03&  15.06& 0.04& S100\\
278.30 &70.3&     &     &  16.97& 0.05 & 15.85& 0.05&  15.37& 0.05&  15.09& 0.04& S100\\
282.31 &74.3&     &     &  17.05& 0.06 & 15.87& 0.08&  15.42& 0.05&  15.06& 0.03&A60\\
283.29 &75.3&     &     &  17.02& 0.04 & 15.83& 0.04&  15.36& 0.04&  15.05& 0.04& S100\\
283.33 &75.3&     &     &       &      & 15.84& 0.04&  15.32& 0.04&  15.02& 0.04& M70\\
284.29 &76.3&     &     &  16.94& 0.05 & 15.84& 0.04&  15.37& 0.04&  15.07& 0.05& S100\\
285.35 &77.4&     &     &  17.10& 0.07 & 15.86& 0.07&  15.42& 0.05&  15.00& 0.05&A60\\
285.37 &77.4&     &     &  17.05& 0.06 & 15.80& 0.03&  15.32& 0.03&       &     & M70\\
288.32 &80.3&     &     &       &      & 15.85& 0.05&  15.32& 0.07&  15.05& 0.11& M70\\
290.32 &82.3&     &     &       &      & 15.87& 0.06&  15.36& 0.06&       &     & M70\\
299.30 &91.3&     &     &  17.15& 0.05 & 15.98& 0.04&  15.48& 0.04&  15.18& 0.04& S100\\
308.44 &100.4&     &     &       &      & 15.97& 0.12&  15.50& 0.05&  15.11& 0.11& S100\\
\hline
\end{longtable}
A60 = 60-cm reflector of Shamakhy Observatory,
C60 = 60-cm reflector of Crimean Observatory of SAI,
C50 = 50-cm meniscus telescope of Crimean Observatory of SAI,
T60 = 60-cm reflector of Star\'a Lesn\'a Observatory,
S100 = 1-m reflector at Simeiz Observatory, 
M70 = 70-cm reflector of Moscow Observatory of SAI,
M20 = 20-cm meniscus telescope of Moscow Observatory of SAI,
K70 = 70-cm reflector of Crimean Astrophysical Observatory (reprocessing
of data by \citet{nazarov2018})
\end{center}

\begin{table}
\begin{center}
\caption{Journal of spectroscopic observations of SN 2018aoq}
\label{tab:spectra}
\begin{tabular}{cccc}
\hline
 JD 2458200+ & Phase &  Exp. time (s) & CCD binning\\
\hline
19.34     & 11.3    &   900  & 8x8 \\
34.36     & 26.4    &  3600  & 4x4 \\
35.33     & 27.3    &  1200  & 8x8 \\
46.35     & 38.4    &  1800  & 8x8 \\
52.45     & 44.5    &  3600  & 4x4 \\
56.40     & 48.4    &  3600  & 4x4 \\
79.39     & 71.4    &  3600  & 4x4 \\
\hline
\end{tabular}
\end{center}
\end{table}

\bsp	
\label{lastpage}
\end{document}